\documentstyle[preprint]{ptptex}
\newcommand{\bra}[1]{\langle {#1} |}     %%
\newcommand{\ket}[1]{| {#1} \rangle}     %%
\newcommand{\bbra}[1]{\langle\!\langle {#1} |}     %%
\newcommand{\kket}[1]{| {#1} \rangle\!\rangle}     %%
\newcommand{\rbra}[1]{( {#1} |}     %%
\newcommand{\rket}[1]{| {#1} )}     %%
\newcommand{\maru}[1]{\breve{#1}} %%
\newcommand{\wtilde}[1]{\widetilde{#1}} %%
\title{%        %You can use \\ for explicit line-break
Two Contrastive Boson-Pair Coherent States\\ 
in Deformed Boson Scheme
}
\author{%       %Use \sc for the family name
Atsushi {\sc Kuriyama}, 
Constan\c{c}a {\sc Provid\^encia}$^{*}$, \\
Jo\~ao da {\sc Provid\^encia}$^{*}$, Yasuhiko {\sc Tsue}$^{**}$ 
and Masatoshi {\sc Yamamura}
}

\inst{%         %Affiliation, neglected when [addenda] or [errata]
Faculty of Engineering, Kansai University, Suita 564-8680, Japan\\
$^{*}$Departamento de Fisica, Universidade de Coimbra, P-3000 Coimbra, 
Portugal\\
$^{**}$Physics Division, Faculty of Science, Kochi University, Kochi 780-8520, 
Japan
}

\recdate{%      %Editorial Office will fill in this.
}

\abst{%       %this abstract is neglected when [addenda] or [errata]
Two types of boson-pair coherent states, which were proposed 
independently by the present authors, are investigated. 
Main conclusion is as follows : Although the two states 
are superposed contrastively 
in terms of the boson-pairs, the expectation values of the boson-pair 
operators for these states are the same as one another with high accuracy. 
}

\begin{document}

\maketitle

%\section{Introduction}

The boson coherent state has played a central role in the studies of many-body 
physics, and for various problems, many ideas for the modification have been 
proposed. The squeezed boson coherent state is one of examples and the present 
authors have studied it extensively.\cite{KPTYs}
Further, recently, from the viewpoint of the $q$-deformed boson scheme, 
the modifications were reinvestigated by the present 
authors.\cite{KPPTYI,KPPTYIII} 
An example is the multiboson coherent state. Under the idea by 
Penson and Solomon,\cite{Penson} the present authors investigated the 
following multiboson coherent state\cite{KPPTYIII} : 
\begin{eqnarray}
& &\kket{c_m(+)}
=\left(\!\!\sqrt{\Gamma_m(+)}\right)^{\!-1}\!\sum_{n=0}^\infty\! 
\sum_{r=0}^{m-1} 
(\!\sqrt{n!}\sqrt{r!})^{-1} \gamma_+^n \delta_+^r 
\left(\!\!\sqrt{C_m(n,r)}\right)^{\!+1}\kket{mn+r} \ , \ \ \ 
\label{1}\\
& &\qquad C_m(n,r)=n!r!\left[(mn+r)!\right]^{-1} \ .
\label{2}
\end{eqnarray}
Here, $\gamma_+$ and $\delta_+$ denote complex parameters and $m$, $n$ and 
$r$ are integers which are shown as follows : 
%\begin{subequations}\label{3}
\vspace{-0.2cm}
\begin{eqnarray}\label{3}
& &m=2, 3, 4, \cdots  \ , \qquad
%\label{3a}\\
%& &
n=0, 1, 2, \cdots  \ , \qquad
%\label{3b}\\
%& &
r=0, 1, 2, \cdots , (m-1)  \ . 
%\label{3c}
\vspace{-0.2cm}
\end{eqnarray}
%\end{subequations}
The quantity $\Gamma_m(+)$ denotes the normalization factor given by 
\vspace{-0.2cm}
\begin{equation}\label{4}
\Gamma_m(+)=\sum_{r=0}^{m-1}(r!)^{-1}(|\delta_+|^2)^r
\left(\sum_{n=0}^\infty (n!)^{-1} C_m(n,r)^{+1}(|\gamma_+|^2)^n\right) \ .
\vspace{-0.3cm}
\end{equation}
The states $\kket{mn+r}$ compose an orthonormal set : 
\begin{eqnarray}
\kket{mn+r}&=&
\bigl(\sqrt{(mn+r)!}\bigl)^{-1}({\wtilde c}^*)^{mn+r} \kket{0} \ , 
\label{5}\\
{\wtilde c}\kket{0}&=&0 \ .
\label{6}
\end{eqnarray}
Here, $({\wtilde c} , {\wtilde c}^*)$ denotes boson operator. 
The condition (\ref{3}) tells us that the set $\{ \kket{mn+r} \}$ 
is complete. Hereafter, the space spanned by the states (\ref{5}) is called 
${\wtilde B}$. We can prove that $\Gamma_m(+)$ is convergent for 
any $m$ in the region 
\begin{equation}\label{7}
|\gamma_+|^2 < \infty \ . \qquad 
(m=2, 3, 4,\cdots) 
\end{equation}
This means that the state $\kket{c_m(+)}$ is normalizable for 
$m=2, 3, 4, \cdots$. It may be clear that the state $\kket{mn+r}$ consists 
of the multiboson $({\wtilde c}^*)^m$. In this sense, we call 
the state $\kket{c_m(+)}$ the multiboson coherent state.

On the other hand, formally, we can set up the following state :
\vspace{-0.2cm}
\begin{eqnarray}\label{8}
\kket{c_m(-)}
=\left(\!\!\sqrt{\Gamma_m(-)}\right)^{\!-1}\!
\sum_{n=0}^\infty \sum_{r=0}^{m-1} 
(\!\sqrt{n!}\sqrt{r!})^{-1} \gamma_-^n \delta_-^r 
\left(\!\!\sqrt{C_m(n,r)}\right)^{\!-1}\!\kket{mn+r} \ . \ \ \ \ 
%\nonumber\\
%& &
\vspace{-0.2cm}
\end{eqnarray}
The normalization factor $\Gamma_m(-)$ is written down as 
\vspace{-0.2cm}
\begin{equation}\label{9}
\Gamma_m(-)=\sum_{r=0}^{m-1}(r!)^{-1}(|\delta_-|^2)^r
\left(\sum_{n=0}^\infty (n!)^{-1} C_m(n,r)^{-1}(|\gamma_-|^2)^n\right) \ .
\vspace{-0.2cm}
\end{equation}
However, $\Gamma_m(-)$ is convergent only for $m=2$ in the region 
\begin{equation}\label{10}
|\gamma_-|^2 < 1/4 \ . \qquad (m=2)
\end{equation}
For the cases $m=3, 4, 5, \cdots$, $\Gamma_m(-)$ is divergent in any 
region of $|\gamma_-|^2$ and in these cases, the state (\ref{8}) 
loses its meaning. The states $\kket{c_{m=2}(\pm)}$ consist of 
the boson-pair $({\wtilde c}^*)^2$ and, then, we call them the 
boson-pair coherent states. We can see that the form of the 
superposition of $\kket{c_2(+)}$ is contrastive to that of $\kket{c_2(-)}$. 
The state $\kket{c_2(-)}$ was already investigated by the present 
authors\cite{KPPYI} with an interesting conclusion : 
With the aid of the state (\ref{8}) with $m=2$, we are able to 
obtain the classical counterpart of the boson-pair, which is 
treated as the set $({\wtilde T}_+ , {\wtilde T}_- , {\wtilde T}_0 )$ 
defined by 
\vspace{-0.2cm}
\begin{equation}\label{11}
{\wtilde T}_+=(1/2) {\wtilde c}^{*2} \ , \quad
{\wtilde T}_-=(1/2) {\wtilde c}^{2} \ , \quad
{\wtilde T}_0=(1/2) {\wtilde c}^{*}{\wtilde c}+1/4 \ .
\end{equation}
The set $({\wtilde T}_+ , {\wtilde T}_- , {\wtilde T}_0)$ obeys the 
$su(1,1)$-algebra. Further, by the requantization, we get the 
Holstein-Primakoff type boson representation for the set 
$({\wtilde T}_+ , {\wtilde T}_- , {\wtilde T}_0)$. 
The above-mentioned scheme presents us a problem : 
To investigate that the state (\ref{1}) for $m=2$ can produce the 
above-mentioned feature. The above task is the purpose of this paper.

By adopting the basic idea of the MYT boson mapping,\cite{MYT} 
we investigate the structure of the states (\ref{1}) and (\ref{8}) 
for $m=2$ from the viewpoint of the deformed boson scheme. 
In order to get the images of $\kket{c_2(\pm)}$, we must prepare 
a new space, which is called, hereafter, ${\hat B}$. 
Following the remarks mentioned in the final part of Ref.\citen{KPPTYIII}, 
we construct the space ${\hat B}$ in terms of the boson 
$({\hat c} , {\hat c}^*)$ and the operator $({\maru d} , {\maru d}^*)$. 
For the boson $({\hat c} , {\hat c}^*)$, we have orthonormal set 
in the form 
\begin{eqnarray}
& &\ket{n}=(\sqrt{n!})^{-1} ({\hat c}^*)^n \ket{0} \ , \qquad
(n=0, 1, 2,\cdots)
\label{12}\\
& &{\hat c}\ket{0}=0 \ . 
\label{13}
\end{eqnarray}
Let the operator $({\maru d} , {\maru d}^*)$ obey the following commutation 
relation : 
\begin{equation}\label{14}
[ {\maru d} , {\maru d}^* ]=1-(m/(m-1)!) ({\maru d}^*)^{m-1}
({\maru d})^{m-1} \ . 
\end{equation}
For the operator $({\maru d} , {\maru d}^*)$, we presuppose the existence 
of the state $\rket{0}$ obeying 
\begin{equation}\label{15}
{\maru d}\rket{0}=0 \ .
\end{equation}
Then, with use of the relations (\ref{14}) and (\ref{15}), 
we are able to obtain the following set:
%\begin{subequations}\label{16}
\vspace{-0.2cm}
\begin{equation}
\rket{r(m)}
=\cases{(\sqrt{r!})^{-1} ({\maru d}^*)^r \rket{0} \ , 
\quad (r=0, 1, 2, \cdots , m-1) 
%\label{16a}
\cr
0 \qquad\qquad\qquad\qquad (r=m, m+1, \cdots) 
%\label{16b}
}\label{16}
\vspace{-0.2cm}
\end{equation}
%\end{subequations}
The operation of ${\maru d}^*$ on the state $\rket{r=m-1(m)}$ 
automatically leads us to 
\begin{equation}\label{17}
{\maru d}^* \rket{r=m-1(m)}=0 \ .
\end{equation}
The relation (\ref{17}) gives us the form in the lower equation of 
(\ref{16}). 
We can also prove the relation 
\begin{equation}\label{18}
{\maru N}_d \rket{r(m)}=r \rket{r(m)}  \ ,\quad
(r=0, 1, 2,\cdots, m-1) \ , \qquad
{\maru N}_d ={\maru d}^*{\maru d} \ .
\end{equation}
Therefore, the set $\{ \rket{r(m)} ; r=0, 1, 2,\cdots , m-1 \}$ 
is orthonormal and complete. 
Then, Dirac's sense,\cite{D} we can set up the following relation 
for $({\maru d} , {\maru d}^*)$ : 
\begin{equation}\label{19}
({\maru d})^m=0 , \qquad ({\maru d}^*)^m=0 \ .
\end{equation}
Combining the states (\ref{12}) with the states (\ref{16}), we 
define the orthonormal set $\{ \ket{n, r(m)} \}$ : 
\begin{equation}\label{20}
\ket{n, r(m)}=\ket{n}\otimes \rket{r(m)} \ .
\end{equation}
The set consisting of the state (\ref{20}) composes the space ${\hat B}$. 
Of course, $\{ \ket{n, r(m)} \}$ is complete. 
Hereafter, we treat the case $m=2$ concretely.

The normalization factors $\Gamma_2(\pm)$ are calculated as follows : 
\begin{subequations}\label{21}
\begin{eqnarray}
& &\Gamma_2(+)=\cosh |\gamma_+|+|\delta_+|^2\cdot |\gamma_+|^{-1}
\sinh |\gamma_+| \ , 
\label{21a}\\
& &\Gamma_2(-)=\bigl(\sqrt{1-4|\gamma_-|^2}\bigl)^{-1}
+|\delta_-|^2\cdot \bigl(\sqrt{1-4|\gamma_-|^2}\bigl)^{-3} \ .
\label{21b}
\end{eqnarray}
\end{subequations}
The images of $\kket{c_2(\pm)}$ in the space ${\hat B}$, which 
we denote $\ket{c_2(\pm)}$, are expressed in the form 
\begin{equation}\label{22}
\ket{c_2(\pm)}={\mib U}\kket{c_2(\pm)} \ .
\end{equation}
Here, ${\mib U}$ is given as 
\begin{equation}\label{23}
{\mib U}=\sum_{n=0}^\infty \sum_{r=0}^1 \ket{n, r(2)}\bbra{2n+r} \ .
\end{equation}
The relation (\ref{22}) presents the following expression 
for $\ket{c_2(\pm)}$ : 
\vspace{-0.1cm}
\begin{eqnarray}\label{24}
& &\ket{c_2(\pm)}=\left(\!\!\sqrt{\Gamma_2(\pm)}\right)^{\!-1}
\!\!\exp\!\left(\!2^{\mp 1}\gamma_\pm{\hat c}^*\left(\!\!
\sqrt{{\hat N}_c+{\maru N}_d+1/2}
\right)^{\!\mp 1}\!\right)
\exp\left(\delta_\pm {\maru d}^*\right)\ket{0} \ . \ \ \ \ 
%\nonumber\\
%& &
\vspace{-0.3cm}
\end{eqnarray}
Here, $\ket{0}$ denotes the vacuum for ${\hat c}$ and ${\maru d}$. 
Since $({\maru d}^*)^2=0$, which comes from the relation (\ref{19}), 
we should note the form 
\begin{equation}\label{25}
\exp\left(\delta_\pm {\maru d}^*\right) \ket{0}
=\ket{0}+\delta_\pm {\maru d}^*\ket{0} \ .
\end{equation}
For the states (\ref{24}), we define the following operators : 
\begin{eqnarray}
& &{\hat \gamma}_\pm = 2^{\pm 1}\bigl(\sqrt{{\hat N}_c+ {\maru N}_d
+1/2}\bigl)^{\pm 1} \ {\hat c} \ , 
\label{26}\\
& &{\maru \delta}_\pm=\bigl(\sqrt{({\hat N}_c+{\maru N}_d+1/2)\cdot
({\maru N}_d+1/2)^{-1}}\bigl)^{\pm 1}\ {\maru d} \ .
\label{27}
\end{eqnarray}
The operators ${\hat \gamma}_\pm$ and ${\maru \delta}_\pm$ satisfy the 
commutation relations 
%\begin{subequations}\label{28}
\begin{eqnarray}\label{28}
& &[ {\hat \gamma}_\pm , {\hat \gamma}_\mp^* ] = 1 \ , \qquad
%\label{28a}\\
%& &
[ {\maru \delta}_\pm , {\hat \gamma}_\mp^* ] = 0 \ , \qquad
%\label{28b}\\
%& &
[ {\hat \gamma}_\pm , {\maru \delta}_\pm ] = 0 \ . 
%\label{28c}
\end{eqnarray}
%\end{subequations}
The operation of ${\hat \gamma}_\pm$ and ${\maru \delta}_\pm$ on the states 
$\ket{c_2(\pm)}$ gives us 
\begin{subequations}\label{29}
\begin{eqnarray}
& &{\hat \gamma}_\pm \ket{c_2(\pm)}=\gamma_\pm\ket{c_2(\pm)} \ , 
\label{29a}\\
& &{\maru \delta}_\pm \ket{c_2(\pm)}=\delta_\pm{\hat P}_1\ket{c_2(\pm)} \ . 
\qquad (\ {\hat P}_1=\rket{0}\rbra{0}\ )
\label{29b}
\end{eqnarray}
\end{subequations}
The relation (\ref{29}) tells us that the states $\ket{c_2(\pm)}$ are 
regarded as coherent states for the operators ${\hat \gamma}_\pm$ and 
${\maru \delta}_\pm$, respectively.

It may be interesting to introduce new parameters 
$(c, c^*)$ and $(d , d^*)$ in our present system, 
which obey the canonicity condition 
\begin{equation}\label{30}
\bra{c_2(\pm)}\partial_c \ket{c_2(\pm)}=c^*/2 \ , 
\qquad
\bra{c_2(\pm)}\partial_d \ket{c_2(\pm)}=d^*/2 \ .
\end{equation}
As was shown in the TDHF theory in canonical form,\cite{MMSK} 
$(c, c^*)$ and $(d, d^*)$ obeying the condition (\ref{30}) can be 
regarded as canonical variables in the boson type. 
The condition (\ref{30}) gives us 
\begin{equation}\label{31}
c\!=\!\gamma_\pm\sqrt{\left(\partial \Gamma_2(\pm)/\partial |\gamma_\pm|^2
\right)
\!\cdot\! \Gamma_2(\pm)^{-1}} \ , \quad
d\!=\!\delta_\pm\sqrt{\left(\partial \Gamma_2(\pm)/\partial |\delta_\pm|^2
\right)
\!\cdot\! \Gamma_2(\pm)^{-1}} \ .
\end{equation}
Since $\Gamma_2(\pm)$ are functions of $|\gamma_\pm|^2$ and $|\delta_\pm|^2$, 
we can express $\gamma_\pm$ and $\delta_\pm$ in terms of $(c,c^*)$ and 
$(d, d^*)$. As was done in Ref.\citen{KPPYI}, the expression for 
$(\gamma_- , \delta_-)$ is easily obtained in the form 
\begin{subequations}\label{32}
\begin{eqnarray}
& &\gamma_-=(1/2)c\bigl(\sqrt{|c|^2+|d|^2+1/2}\bigl)^{-1}\ , 
\label{32a}\\
& &\delta_-=d\sqrt{(1/2+|d|^2)(1-|d|^2)^{-1}}
\bigl(\sqrt{|c|^2+|d|^2+1/2}\bigl)^{-1} \ .
\label{32b}
\end{eqnarray}
\end{subequations}
Of course, we used the relation (\ref{21b}). In the case of 
$(\gamma_+ , \delta_+)$, the relations (\ref{21a}) and (\ref{31}) give 
us 
\begin{subequations}\label{33}
\begin{eqnarray}
& &|\gamma_+|^2=(1-|d|^2)^{-1}\cdot
\left(2|c|^2+|d|^2(1-h(|\gamma_+|))\right) h(|\gamma_+|) \ , 
\label{33a}\\
& &|\delta_+|^2=|d|^2(1-|d|^2)^{-1}\cdot h(|\gamma_+|) \ , 
\label{33b}
\end{eqnarray}
\end{subequations}
\vspace{-0.6cm}
\begin{equation}\label{34}
h(|\gamma_+|)=|\gamma_+|\coth |\gamma_+| \ .\qquad\qquad\qquad\qquad\qquad
\qquad\qquad
\end{equation}
As is clear from the form (\ref{33}), it may be impossible to 
have simple expression for $(|\gamma_+|^2 , |\delta_+|^2)$ in terms 
of $(|c|^2 , |d|^2)$. Therefore, we try to give approximate expressions 
in the two extreme regions : $|\gamma_+|\rightarrow 0$ and 
$|\gamma_+|\rightarrow \infty$. First, we note that 
the following expressions can be derived : 
\begin{eqnarray}
& &h(|\gamma_+|)=1\!+\!(1/3)|\gamma_+|^2\!-\!(1/45)|\gamma_+|^4
\!+\!(2/945)|\gamma_+|^6- \cdots \ , \ \ \hbox{\rm for}\quad
|\gamma_+|<\pi \ , \ \ \ \ \ \ \ 
%\nonumber\\
%& &
\label{35}\\
& &h(|\gamma_+|)=|\gamma_+|(1+2e^{-2|\gamma_+|}+2e^{-4|\gamma_+|}
+2e^{-6|\gamma_+|} + \cdots ) \ , \quad \hbox{\rm for}\quad
|\gamma_+| \ge 0 \ . 
\label{36}
\end{eqnarray}
The forms (\ref{35}) and (\ref{36}) are useful for the regions 
$|\gamma_+|^2\rightarrow 0$ and $|\gamma_+|^2\rightarrow \infty$, 
respectively. 
By using the relation (\ref{35}), the form (\ref{33a}) is reduced to 
the following for the region $|\gamma_+|\rightarrow 0$ : 
\begin{eqnarray}\label{37}
& &|\gamma_+|^2=2|c|^2+(2/3)(|c|^2+|d|^2)|\gamma_+|^2 
-(2/45)(|c|^2+2|d|^2)|\gamma_+|^4+\cdots \ .\ \ 
\end{eqnarray}
The relation (\ref{37}) leads us to the approximate expressions 
for $(\gamma_+ , \delta_+)$ in the region $|\gamma_+|\rightarrow 0$ : 
%\begin{subequations}\label{38}
\begin{eqnarray}\label{38}
& &\gamma_+\!=\!2c\sqrt{(1/3)(|c|^2\!+\!|d|^2)\!+1/2} \ , \quad
%\label{38a}\\
%& &
\delta_+\!=\!d\bigl(\!\sqrt{1-|d|^2}\bigl)^{-1}\!\sqrt{(2/3)|c|^2+1} \ .
%\label{38b}
\end{eqnarray}
%\end{subequations}
The relation (\ref{36}) gives us the following form which is 
reduced from the relation (\ref{33}) for the region 
$|\gamma_+|\rightarrow \infty$ : 
\begin{eqnarray}\label{39}
|\gamma_+|^2&=&(2|c|^2+|d|^2)^2
+4\left((2|c|^2+|d|^2)^2-2|d|^2(2|c|^2+|d|^2)|\gamma_+|\right) e^{-2|\gamma_+|}
\nonumber\\
&+&8\left((2|c|^2+|d|^2)^2-4|d|^2(2|c|^2+|d|^2)|\gamma_+|+2|d|^4
|\gamma_+|^2\right)e^{-4|\gamma_+|}+\cdots \ .\qquad \ \ 
\end{eqnarray}
Then, we have the approximate expressions for $(\gamma_+ , \delta_+)$ 
in the region $|\gamma_+|\rightarrow \infty$ : 
%\begin{subequations}\label{40}
\begin{eqnarray}\label{40}
& &\gamma_+=2c\sqrt{|c|^2+|d|^2+|d|^4/4|c|^2} \ , \quad
%\label{40a}\\
%& &
\delta_+=d\bigl(\!\sqrt{1-|d|^2}\bigl)^{-1}\!\sqrt{2|c|^2+|d|^2} \ .\ \ 
%\label{40b}
\end{eqnarray}
%\end{subequations}
Since $\ket{c_2(+)}$ is different from $\ket{c_2(-)}$, it is meaningless 
to compare the results (\ref{38}) and (\ref{40}) with (\ref{32}) directly.

The comparison of both results can be done through the expectation 
values of the images of ${\wtilde T}_{\pm, 0}$, defined in the relation 
(\ref{11}). The operators ${\hat {\mib T}}_{\pm, 0}$ defined by 
${\mib U}{\wtilde T}_{\pm, 0}{\mib U}^\dagger$ are given by 
\begin{eqnarray}\label{41}
& &{\hat {\mib T}}_{\!+}\!
=\!{\hat c}^*\sqrt{{\hat N}_c\!+\!{\maru N}_d\!+\!1/2} 
\ , 
\ \ \ 
%\nonumber\\
%& &
{\hat {\mib T}}_{\!-}\!=\!\sqrt{{\hat N}_c\!+\!{\maru N}_d\!+\!1/2} \ {\hat c}\ , 
\ \ \ 
%\nonumber\\
%& &
{\hat {\mib T}}\!_0\!=\!{\hat N}_c\!+\!{\maru N}_d/2\!+\!1/4 \ . \ \ \ \ \ \ 
\end{eqnarray}
The expectation value of ${\hat {\mib T}}\!_+$ for $\ket{c_2(-)}$ 
is shown in Ref.\citen{KPPYI} : 
\begin{equation}\label{42}
\bra{c_2(-)}{\hat {\mib T}}\!_+\ket{c_2(-)}
=c^*\sqrt{|c|^2+|d|^2+1/2} \ .
\end{equation}
In the case of $\ket{c_2(+)}$, we note the relation ${\hat {\mib T}}\!_+
={\hat \gamma}_+^*/2$. 
Then, in the region $|\gamma_+|\rightarrow 0$, i.e., $|c|\rightarrow 0$, 
the first equation in (\ref{38}) gives us 
\begin{subequations}
\begin{equation}\label{43a}
\bra{c_2(+)}{\hat {\mib T}}\!_+\ket{c_2(+)}
=c^*\sqrt{(1/3)(|c|^2+|d|^2)+1/2} \ .
\end{equation}
In the region $|\gamma_+|\rightarrow \infty$, i.e., $|c|\rightarrow \infty$, 
the first equation in (\ref{40}) leads us to 
\begin{equation}\label{43b}
\bra{c_2(+)}{\hat {\mib T}}\!_+\ket{c_2(+)}
=c^*\sqrt{|c|^2+|d|^2+|d|^4/4|c|^2} \ .
\end{equation}
\end{subequations}
For ${\hat {\mib T}}\!_0$, both forms $\ket{c_2(\pm)}$ 
give us the common result : 
\begin{equation}\label{44}
\bra{c_2(\pm)}{\hat {\mib T}}\!_0\ket{c_2(\pm)}
=|c|^2+|d|^2/2+1/4 \ .
\end{equation}
The operator ${\hat {\mib c}}^*$ defined by 
${\mib U}{\wtilde c}^*{\mib U}^\dagger$ is also interesting. 
It is given by 
\begin{equation}\label{45}
{\hat {\mib c}}^*=\sqrt{2}
\bigl({\hat c}^*{\maru d}+{\maru d}^*\sqrt{{\hat c}^*{\hat c}
+{\maru d}^*{\maru d}+1/2}\bigl) \ .
\end{equation}
The expectation value for $\ket{c_2(-)}$ is calculated as 
\begin{equation}\label{46}
\bra{c_2(-)}{\hat {\mib c}}^*\ket{c_2(-)}
=\!\sqrt{(1-|d|^2)(1/2+|d|^2)^{-1}}
\bigl(\!c^*d+d^*\sqrt{|c|^2+|d|^2+1/2}\bigl)\ . 
\end{equation}
The above is shown in Ref.\citen{KPPYI}. 
The expectation value for $\ket{c_2(+)}$ is given in the form 
\vspace{-0.2cm}
\begin{equation}\label{47}
\bra{c_2(+)}{\hat {\mib c}}^*\ket{c_2(+)}
=(\gamma_+^*\delta_+ +\delta_+^*h(|\gamma_+|))
(|\delta_+|^2+h(|\gamma_+|))^{-1} \ .
\vspace{-0.0cm}
\end{equation}
In the region $|\gamma_+|\rightarrow 0$, i.e., $|c|\rightarrow 0$, the 
relation (\ref{47}) is approximated as 
%\vspace{-0.2cm}
\begin{subequations}\label{48}
\begin{eqnarray}\label{48a}
\bra{c_2(+)}{\hat {\mib c}}^*\ket{c_2(+)}
&=&\sqrt{(1-|d|^2)(1/2)^{-1}}
\bigl(c^*d\sqrt{(1+(2/3)(|c|^2+|d|^2)(1+(2/3)|c|^2)^{-1}}
\nonumber\\
& &\qquad\qquad\qquad\qquad\qquad\qquad\qquad
+d^*\sqrt{(1/3)|c|^2+1/2}\bigl)\ .
\vspace{-0.2cm}
\end{eqnarray}
Further, the approximate form in the region $|\gamma_+|\rightarrow \infty$, 
i.e., $|c|\rightarrow \infty$, is obtained as follows : 
\vspace{-0.2cm}
\begin{equation}\label{48b}
\bra{c_2(+)}{\hat {\mib c}}^*\ket{c_2(+)}
=\sqrt{(1-|d|^2)(1/2)^{-1}}
\bigl(c^*d+d^*\sqrt{|c|^2+|d|^2/2}\bigl)\ .
\end{equation}
\vspace{-0.2cm}
\end{subequations}

\vspace{-0.2cm}
The expectation values of ${\hat {\mib T}}\!_+$ and ${\hat {\mib c}}^*$ shown 
in the relations (\ref{42}) and (\ref{46}), respectively, for 
the state $\ket{c_2(-)}$ are exactly calculated in any region of 
$|c|^2$ and the case of ${\hat {\mib T}}\!_0$ shown in the form (\ref{44}) 
is also in the same situation as the above. Further, the expectation 
values of ${\hat {\mib T}}_{\pm, 0}$ give us the Poisson brackets 
under the form which are completely the same as those of the 
commutation relations for ${\wtilde T}_{\pm, 0}$ in Dirac's sense. 
This fact was shown in Ref.\citen{KPPYI}. 
From the above reason, we can see that the state $\ket{c_2(-)}$ 
gives us the classical counterpart for the set 
$({\wtilde T}_+ , {\wtilde T}_- , {\wtilde T}_0)$. 
On the other hand, we gave the approximate expressions in the two 
extreme regions for the state $\ket{c_2(+)}$. 
However, the leading terms with respect to $|c|^2$ are exact and, 
the forms are almost the same as those in the case of $\ket{c_2(-)}$ 
in the framework of the leading term. 
Therefore, we can able to reproduce the same aspects as those for 
$\ket{c_2(-)}$ with rather high accuracy in spite of contrastive 
superpositions for the two boson-pair coherent states. 
Further, it can be expected that the approximation shown in the 
relation (38) in Ref.\citen{KPPTYIII} may be permitted. 
We note again that the state $\ket{c_2(+)}$ can be extended to the case 
$m=3, 4, 5, \cdots$, but the state $\ket{c_2(-)}$ cannot be done. 
From the above reason, $\ket{c_m(+)}$ may be 
superior to $\ket{c_m(-)}$. This is our conclusion. It may be interesting 
to apply the state $\ket{c_2(+)}$ to the Lipkin model, the prototype 
of which was already presented in Ref.\citen{KPPYII}.

%\section*{Acknowledgements}

\end{document}